\documentclass[prb,twocolumn,aps,amssymb,footinbib]{revtex4}

\usepackage{amssymb,amsfonts,amsmath}
\usepackage{graphicx}

\usepackage{times}
\usepackage{color}
\usepackage{bm}
\usepackage{xr}

\newcommand{\qv}{\mathbf{q}}
\newcommand{\kv}{\mathbf{k}}

\begin{document}

\title{Topological Boundary Modes in Isostatic Lattices}

\author{C. L. Kane and T. C. Lubensky}
\affiliation{Dept. of Physics and Astronomy, University of
Pennsylvania, Philadelphia, PA 19104}

\begin{abstract}

Frames, or lattices consisting of mass points connected by
rigid bonds or central-force springs, are important model
constructs that have applications in such diverse fields as
structural engineering, architecture, and materials science.
The difference between the number of bonds and the number of
degrees of freedom of these lattices determines the number of
their zero-frequency ``floppy modes". When these are balanced,
the system is on the verge of mechanical instability and is
termed isostatic. It has recently been shown that certain
extended isostatic lattices exhibit floppy modes localized at
their boundary. These boundary modes are insensitive to local
perturbations, and appear to have a topological origin,
reminiscent of the protected electronic boundary modes that
occur in the quantum Hall effect and in topological insulators.
In this paper we establish the connection between the
topological mechanical modes and the topological band theory of
electronic systems, and we predict the existence of new
topological bulk mechanical phases with distinct boundary
modes.  We introduce model systems in one and two dimensions
that exemplify this phenomenon.

\end{abstract}

\maketitle

Isostatic lattices provide a useful reference point for
understanding the properties of a wide range of systems on the
verge of mechanical instability, including network glasses
\cite{Phillips1981,Thorpe1983}, randomly diluted lattices near
the rigidity percolation threshold
\cite{FengSen1984,JacobsTho1995}, randomly packed particles
near their jamming threshold
\cite{LiuNag1998,LiuNag2010a,LiuNag2010b,TorquatoSti2010,Wyartwit2005b,Wyart2005},
and biopolymer networks
\cite{WilhelmFre2003,HeussingerFrey2006,HuismanLub2011,BroederszMac2011}.
There are many periodic lattices, including the square and
kagome lattices in $d=2$ dimensions and the cubic and
pyrochlore lattices in $d=3$, that are locally isostatic with
coordination number $z=2d$ for every site under periodic
boundary conditions. These lattices, which are the subject of
this paper, have a surprisingly rich range of elastic responses
and phonon structures
\cite{SouslovLub2009,MaoLub2010,MaoLub2011a,MaoLub2013b,KapkoGue2009}
that exhibit different behaviors as bending forces or
additional bonds are added.

The analysis of such systems dates to an 1864 paper by James
Clerk Maxwell \cite{ Maxwell1864} that argued that a lattice
with $N_s$ mass points and $N_b$ bonds has $N_0 = d N_s
- N_b$ zero modes.   Maxwell's count is incomplete, though,
because $N_0$ can exceed $d N_s-N_b$ if there are
$N_{\rm ss}$ states of self-stress, where springs can be under
tension or compression with no net forces on the masses.  This
occurs, for example, when masses are connected by straight
lines of bonds under periodic boundary conditions.  A more
general Maxwell relation \cite{Calladine1978},
\begin{equation}
\nu \equiv N_0 - N_{\rm ss} = d N_s - N_b ,
\label{eq1}
\end{equation}
is valid for infinitesimal distortions.

In a locally isostatic system with periodic boundary
conditions, $N_0=N_{\rm ss}$.  The square and kagome lattices
have one state of self-stress per straight line of bonds and
associated zero modes along lines in momentum space. Cutting a
section of $N$ sites from these lattices removes states of
self-stress and ${\cal O}(\sqrt{N})$ bonds and necessarily
leads to ${\cal O}(\sqrt{N})$ zero modes, which are essentially
identical to the bulk zero modes. Recently Sun \emph{et al.}
\cite{SunLub2012} studied a twisted kagome lattice in which
states of self-stress are removed by rotating adjacent
site-sharing triangles in opposite directions.  This simple
modification converts lines of zero modes in the
untwisted lattice to gapped modes of nonzero frequency (except
for $\qv=0$) and localizes the required zero modes in the cut
lattice to its surfaces.

\begin{figure}
\includegraphics[width=3.in]{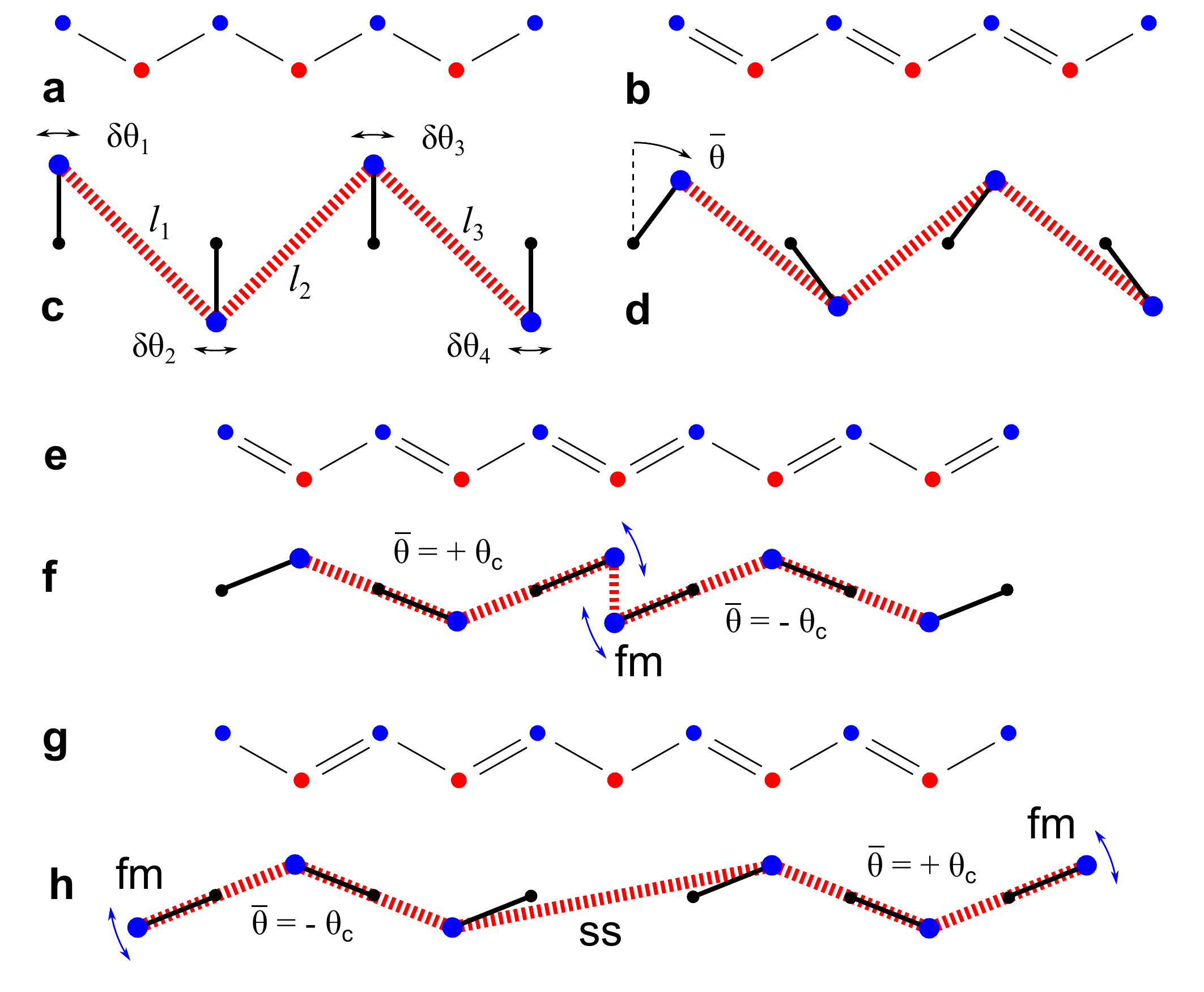}
\caption{{\bf One-dimensional SSH and isostatic lattice models.} (a) and (b) depict
the SSH model of polyacetalene, with A and B sublattices indicated in blue and red.
(a) describes the gapless state with all bonds identical, while (b) describes the
gapped AB dimerized state, with double (single) bonds on the AB (BA) bonds. The BA
dimerized state with single and double bonds interchanged is not shown.
(c) and (d) show the 1D isostatic lattice model in which masses, represented the
larger blue dots are connected by springs in red and are constrained
to rotate about fixed pivot points represented by small black dots.
(c) is the gapless
high-symmetry state with $\bar\theta=0$, and (d) is the gapped lower-symmetry phase
with  $\bar\theta >0$.   (c) and (d) are equivalent to (a) and (b) if we identify the masses (springs) with the A (B) sublattice sites.
(e) shows a domain wall in polyacetalene connecting the AB and BA dimerized states.
There is a topologically protected zero-energy state associated with the A sublattice at the defect. (f) shows
the equivalent state for the isostatic model with a topologically protected
zero-frequency mode at the domain wall. (g) shows domain wall connecting the BA and AB dimerized states, which has a zero energy state associated with the B sublattice.
(h) shows the equivalent isostatic state with a state of self-stress at the domain wall.}
\label{Fig1}
\end{figure}

These boundary zero modes are robust and insensitive to local
perturbations.   Boundary modes also occur in electronic
systems, such as the quantum Hall effect
\cite{halperin82,haldane88} and topological insulators
\cite{km05b,bhz06,mb07,fkm07,HasanKane2010,QiZhang2011}. In
this paper we establish the connection between these two
phenomena. Our analysis allows us to predict the existence of
new topologically distinct bulk mechanical phases and to
characterize the protected modes that occur on their boundary.
We introduce a 1D model that illustrates this phenomenon in its
simplest form and maps directly to the Su-Schrieffer-Heeger
(SSH) model \cite{ssh} for the electronic excitations
of polyacetylene ($(CH_2)_n$), a linear polymer with
alternating single and double bonds between carbon atoms as
show in figure \ref{Fig1} that has toplogically protected
electron states at free ends and at interfaces. We then prove
an index theorem that generalizes equation (\ref{eq1}) and
relates the local count of zero modes on the boundary to the
topological structure of the bulk. We introduce a deformed
version of the kagome lattice model that exhibits distinct
topological phases. The predictions of an index theorem are
verified explicitly by solving for the boundary modes in this
model.  Finally, we show that some of the distinctive features
of the topological phases can be understood within a continuum
elastic theory.

\section*{Mechanical Modes and Topological Band Theory}

A mechanical system of masses $M$ connected by springs with
spring constant $K$ is characterized by its equilibrium matrix
\cite{Calladine1978} $Q$, which relates forces $F_i = Q_{im}
T_m$ to spring tensions $T_m$.   $i$ labels the $d N_s$ force
components on the $N_s$ sites, and $m$ labels the $N_b$ bonds.
Equivalently, $e_m = Q^T_{mi} u_i$ relates bond extensions
$e_m$ to site displacements $u_i$. The squared normal mode
frequencies $\omega^2_n$ are eigenvalues of the dynamical
matrix $D = Q Q^T$, where we set $K/M$ to unity. Displacements
$u_i$ that do not lead to stretched bonds satisfy $Q^T u_i= 0$
and define the null space of $Q^T$, or equivalently its kernel
${\rm ker}\ Q^T$.  The dimension of this null space $N_0 = {\rm
dim}\ {\rm ker}\ Q^T$ gives the number of independent zero
modes.  Similarly, the null space of $Q$ gives the $N_{\rm ss}
= {\rm dim}\ {\rm ker}\ Q $ states of self-stress. The global
counts of these two kinds of zero modes are related by the
rank-nullity theorem \cite{ Calladine1978}, which may be
expressed as an index theorem \cite{nakahara}. The index of
$Q$, defined as $\nu = {\rm dim}\ {\rm ker}\ Q^T - {\rm dim}\
{\rm ker}\ Q$, is equal to the difference between the number of
rows and columns of $Q$, and is given by equation (\ref{eq1}).

At first sight, the mechanical problem and the quantum
electronic problem appear different.  Newton's laws are
second-order equations in time, while the Schrodinger equation
is first order.  The eigenvalues of $D$ are positive definite,
while an electronic Hamiltonian has positive and negative
eigenvalues for the conduction and valence bands. To uncover
the connection between the two problems we draw our inspiration
from Dirac, who famously took the ``square root" of the Klein
Gordon equation \cite{dirac28}. To take the square root of the
dynamical matrix, note that $D = QQ^T$ has a supersymmetric
partner $\tilde D = Q^TQ$ with the same non-zero eigenvalues.
 Combining $D$ and $\tilde D$
gives a matrix that is a perfect square,
\begin{equation}
{\cal H} = \left(\begin{array}{cc} 0 & Q \\ Q^T & 0 \end{array}\right); \quad
{\cal H}^2 = \left(\begin{array}{cc} Q Q^T & 0 \\ 0 & Q^T Q \end{array}\right).
\label{eq2}
\end{equation}
The spectrum of ${\cal H}$ is identical to that of $D$, except
for the zero modes.   Unlike $D$, the zero modes of ${\cal H}$
include {\it both} the zero modes of $Q^T$ and $Q$, which are
eigenstates of $\tau^z = {\rm diag}(1_{dN_s},-1_{N_b})$
distinguished by their eigenvalue, $\pm 1$. (While the concept
of supersymmetry is not essential for deriving this result,
there is an interesting connection between the analysis leading
to equation (\ref{eq2}) and the theory of supersymmetric
quantum mechanics\cite{Witten81,susyqm}.)

Viewed as a Hamiltonian, $\cal H$ can be analyzed in the
framework of topological band theory \cite{HasanKane2010}.  It
has an intrinsic ``particle-hole" symmetry, $\{{\cal H},\tau^z
\} = 0$, that guarantees eigenstates come in $\pm E$ pairs.
Since $Q_{im}$ is real, ${\cal H} = {\cal H}^*$ has
``time-reversal" symmetry. These define symmetry class BDI
\cite{schnyder08}.  In one-dimension, gapped Hamiltonians in
this class, such as the SSH model, are characterized by an
integer topological invariant $n \in \mathbb{Z}$ that is a
property of the Bloch Hamiltonian ${\cal H}(k)$ (or
equivalently $Q(k)$) defined at each wavenumber $k$ in the
Brillouin zone (BZ). A mapping of ${\cal H}(k)$ to the complex
plane is provided by $\det Q(k) = |\det Q(k)|e^{i \phi(k)}$. If
bulk modes are all gapped then $|\det Q(k)|$ is nonzero and
$Q(k) \in GL_p$, where $p$ is the dimension of $Q(k)$.  $Q(k)$
is then classified by the integer winding number of $\phi(k)$
around the BZ: ($\phi(k+G)-\phi(k) =2 \pi n$, where $G$ is a
reciprocal lattice vector), which defines an element of the
homotopy group $\pi_1(GL_p) = \mathbb{Z}$. A consequence is
that a domain wall across which $n$ changes is associated with
topologically protected zero modes
\cite{ssh,jackiw76,volovik03}.  Below, we present an index
theorem that unifies this bulk-boundary correspondence with
equation (\ref{eq1}) and shows how it can be applied to
$d$-dimensional lattices, which form the analog of weak
topological insulators \cite{fkm07}.

Topological edge modes have been previously predicted in 2D
photonic\cite{haldane08,wang08} and phononic\cite{prodan09}
systems.  These differ from the present theory because they
occur in systems with bandgaps at finite frequency and broken
time-reversal symmetry (symmetry class A). Localized end modes
were found in a time-reversal invariant 1D model (class AI)
\cite{Berg11}.  However, the presence of those finite frequency
modes is not topologically guaranteed.

\section*{One-Dimensional Model}

Before discussing the index theorem we introduce a simple 1D
elastic model, equivalent to the SSH model, that illustrates
the topological modes in their simplest setting. Consider a 1D
system of springs connecting masses constrained to rotate at a
radius $R$ about fixed pivot points. In Fig.~1c the spring
lengths are set so that the equilibrium configuration is
$\langle \theta_i\rangle = 0$. Fig.~1d shows a configuration
with shorter springs with $\langle \theta_i\rangle =
\bar\theta$. Expanding in deviations $\delta\theta_i$ about
$\bar\theta$, the extension of spring $m$ is $\delta\ell_m =
Q^T_{mi} \delta\theta_i$, with $Q^T_{mi} = q_1(\bar\theta)
\delta_{m,i} + q_2(\bar\theta)\delta_{m,i+1}$ and $q_{1(2)} = r
\cos\bar\theta( r\sin\bar\theta \pm
1)/\sqrt{4r^2\cos^2\bar\theta+1}$.    The normal mode
dispersion is $\omega(k) = |Q(k)|$, where $Q(k) = q_1 + q_2
e^{ik}$.  When $\bar\theta=0$, $q_1=-q_2$, and there are
gapless bulk modes near $k=0$.  For a finite system with $N$
sites and $N-1$ springs, there are no states of self stress and
only a single extended zero mode, as required by equation
(\ref{eq1}). For $\bar\theta\ne 0$ the bulk spectrum has a gap.
In this case, the zero mode required by equation (\ref{eq1}) is
localized at one end or the other, depending on the sign of
$\bar\theta$. The $\bar\theta>0$ and $\bar\theta<0$ phases are
topologically distinct in the sense that it is impossible to
tune between the two phases without passing through a
transition where the gap vanishes.  The topological distinction
is captured by the winding number of the phase of $Q(k)$, which
is $1$ ($0$) for $|q_1| <(>) |q_2|$.

Viewed as a quantum Hamiltonian, equation (\ref{eq2}) for this
model is identical to the SSH model \cite{ssh}, as indicated in
Fig.~1(a,b).  The sites and the bonds correspond, respectively,
to the A and B sublattices of the SSH model. For $\bar\theta=0$
the bonds in the SSH model are the same (Fig.~1a), while for
$\bar\theta\ne 0$ they are dimerized (Fig.~1b).  The two
topological phases correspond to the two dimerization patterns
for polyacetalene.  As is well known for the SSH model
\cite{ssh,jackiw76}, an interface between the two dimerizations
binds a zero mode, as indicated in Fig.~1(e,g).   This is most easily seen for $\bar\theta =
\pm \theta_c$ where the springs are colinear with the bars, so
that $q_1$ or $q_2 = 0$. Fig.~1f shows a domain wall between
$+\theta_c$ and $-\theta_c$, in which the center two sites
share a localized floppy mode. Fig.~1h shows an interface
between $-\theta_c$ and $+\theta_c$ with a state of self-stress
localized to the middle three bonds, in addition to floppy
modes localized at either end. As long as there is a bulk gap,
the zero modes cannot disappear when $\bar\theta$ deviates from
$\pm \theta_c$.  The zero modes remain exponentially localized,
with a localization length that diverges when
$\bar\theta\rightarrow 0$.

\section*{Index Theorem}

There appear to be two distinct origins for zero modes.   In
equation (\ref{eq1}) they arise because of a mismatch between
the number of sites and bonds, while at a domain wall they arise
in a location where there is no local mismatch.  To unify them, we
generalize the index theorem so that it determines the
zero-mode count $\nu^S$ in a subsystem $S$ of a larger system.
This is well defined provided the boundary of $S$ is deep in a
gapped phase where zero modes are absent.  We will show there
are two contributions,
\begin{equation}
\nu^S = \nu^S_L + \nu^S_T ,
\label{eq3}
\end{equation}
where $\nu_L^S$ is a local count of sites and bonds in $S$ and
$\nu^S_T$ is a topological count, which depends on the
topological structure of the gapped phases on the boundary of
$S$.

To prove equation (\ref{eq3}) and to derive formulas for
$\nu^S_T$ and $\nu^S_L$, we adapt a local version of the index
theorem originally introduced by Callias \cite{callias,bott,
hirayama,niemi} to allow for the possibility of non-zero
$\nu^S_L$. The details of the proof are given in the
supplementary material.  Here we will focus on the results.
Consider a $d$-dimensional system described by a Hamiltonian
matrix ${\cal H}_{\alpha\beta}$, where $\alpha$
labels a site or a bond centered on ${\bf r}_\alpha$. The count
of zero modes in $S$ may be written
\begin{equation}
\nu^S = \lim_{\epsilon\rightarrow 0} {\rm Tr}\left[\tau^z\rho_S(\hat{\bf r}) {i\epsilon\over{{\cal H} + i\epsilon}} \right] ,
\label{eq4}
\end{equation}
where $\hat {\bf r}_{\alpha\beta} = \delta_{\alpha\beta} {\bf
r}_\alpha$. The region $S$ is defined by the support function
$\rho^S({\bf r}) = 1$ for ${\bf r} \in S$ and $0$ otherwise. It
is useful to consider $\rho^S({\bf r})$ to vary smoothly
between 1 and 0 on the boundary $\partial S$.   Expanding the
trace in terms of eigenstates of ${\cal H}$ shows that
only zero modes with support in $S$ contribute.

In the supplementary material we show that equation (\ref{eq4})
can be rewritten as equation (\ref{eq3}) with
\begin{equation}
\nu^S_L = {\rm Tr}\left[\rho^S(\hat{\bf r})\tau^z \right]
\label{nul}
\end{equation}
and
\begin{equation}
\nu^S_T = \int_{\partial S} {d^{d-1}S\over V_{\rm cell}}  {\bf R}_T \cdot \hat n ,
\label{nut}
\end{equation}
where the integral is over the boundary of $S$ with inward
pointing normal $\hat n$.  ${\bf R}_T= \sum_i n_i {\bf a}_i$ is
a Bravais lattice vector characterizing the periodic crystal in
the boundary region that can be written in terms of primitive
vectors ${\bf a}_i$ and integers
\begin{equation}
n_i = {1\over{2\pi i}}\oint_{C_i} d{\bf k}\cdot {\rm Tr}[Q({\bf k})^{-1}
\nabla_{\bf k} Q({\bf k}) ].
= {1\over{2\pi }}\oint_{C_i} d{\bf k}\cdot \nabla_{\bf k} \phi ({\bf k}) .
\label{ni}
\end{equation}
Here $C_i$ is a cycle of the BZ connecting ${\bf k}$ and ${\bf
k} + {\bf b}_i$, where ${\bf b}_i$ is a primitive reciprocal
vector satisfying ${\bf a}_i \cdot {\bf b}_j = 2\pi
\delta_{ij}$.  $n_i$ are winding numbers of the phase of $
\det Q({\bf k})$ around the cycles of the BZ, where $Q({\bf
k})$ is the equilibrium matrix in a Bloch basis.

To apply equations (\ref{nut}) and (\ref{ni}), it is important
that the winding number be independent of path.  This is the
case if there is a gap in the spectrum.  We will also apply
this when the gap vanishes for acoustic modes at ${\bf k}=0$.
This is okay because the acoustic mode is not topological in
the sense that it can be gapped by a weak
translational-symmetry-breaking perturbation.  This
means the winding number is independent of ${\bf k}$ even near
${\bf k}=0$. It is possible, however, that there can be
topologically protected gapless points.  These would be point
zeros around which the phase of $\det Q({\bf k})$ advances by
$2\pi$. These lead to topologically protected bulk modes that
form the analog of a ``Dirac semimetal" in electronic systems
like graphene. These singularities could be of interest, but
they do not occur in the model we study below.

A second caveat for equation (\ref{ni}) is that, in general,
the winding number is not gauge invariant and depends on how
the sites and bonds are assigned to unit cells.   In the
supplementary material we show that for an
isostatic lattice with uniform coordination it is possible to adopt a ``standard
unit cell" with basis vectors ${\bf d}_{i(m)}$ for the $n_s$
sites ($d n_s$ bonds) per cell that satisfy ${\bf r}_0 = d
\sum_i {\bf d}_i - \sum_m {\bf d}_m = 0$.   $Q({\bf k})$ is
defined using Bloch basis states $|{\bf k},a=i,m\rangle \propto
\sum_{{\bf R}}\exp i {\bf k}\cdot({\bf R} + {\bf d}_a)|{\bf R}
+ {\bf d}_a \rangle$, where ${\bf R}$ is a Bravais lattice
vector.   In this gauge, ${\bf R}_T$ is uniquely
defined and the zero-mode count is given by equations
(\ref{eq3}) and (\ref{nul})-(\ref{ni}).

To determine the number of zero modes per unit cell on an edge
indexed by a reciprocal lattice vector ${\bf G}$, consider a
cylinder with axis parallel
to ${\bf G}$ and define the region $S$ to be the points nearest
to one end of the cylinder (See Supplementary Fig.~1).
$\nu^S_T$ is determined by evaluating equation (\ref{nut}),
with the aid of equation (\ref{ni}) on $\partial S$ deep in the
bulk of the cylinder. It follows that
\begin{equation}
\tilde\nu_T \equiv \nu^S_T/N_{\rm cell} = {\bf G} \cdot {\bf R}_T/2\pi.
\label{nutcell}
\end{equation}
The local count, $\nu^S_L$, depends on the details of the
termination at the surface and can be determined by evaluating
the macroscopic ``surface charge" that arises when charges $+d$
($-1$) are placed on the sites (bonds) in a manner analogous to
the ``pebble game" \cite{JacobsTho1995}. This can be found by
defining a bulk unit cell with basis vectors $\tilde {\bf d}_a$
that accommodates the surface with no leftover sites or bonds
(see Fig.~4a below).   Note that this unit cell depends on the
surface termination and, in general, will be different from the
``standard" unit cell used for $\nu^S_T$.   The local count is
then the surface polarization charge given by the dipole moment
per unit cell.   We find
\begin{equation}
\tilde\nu_L \equiv \nu^S_L/N_{\rm cell} = {\bf G}\cdot {\bf R}_L /2\pi,
\label{nulcell}
\end{equation}
where
\begin{equation}
{\bf R}_L = d \sum_{{\rm sites }\ i} \tilde{\bf d}_i - \sum_{{\rm bonds }\ m} \tilde {\bf d}_m.
\label{rl}
\end{equation}
${\bf R}_L$ is similar to ${\bf r}_0$ defined above (which is
assumed to be zero), but it is in general a different Bravais
lattice vector.    The total zero mode count on the surface
then follows from equations (\ref{eq3}), (\ref{nutcell}), and
(\ref{nulcell}).

\section*{Deformed Kagome Lattice Model}

\begin{figure}
\includegraphics[width=3in]{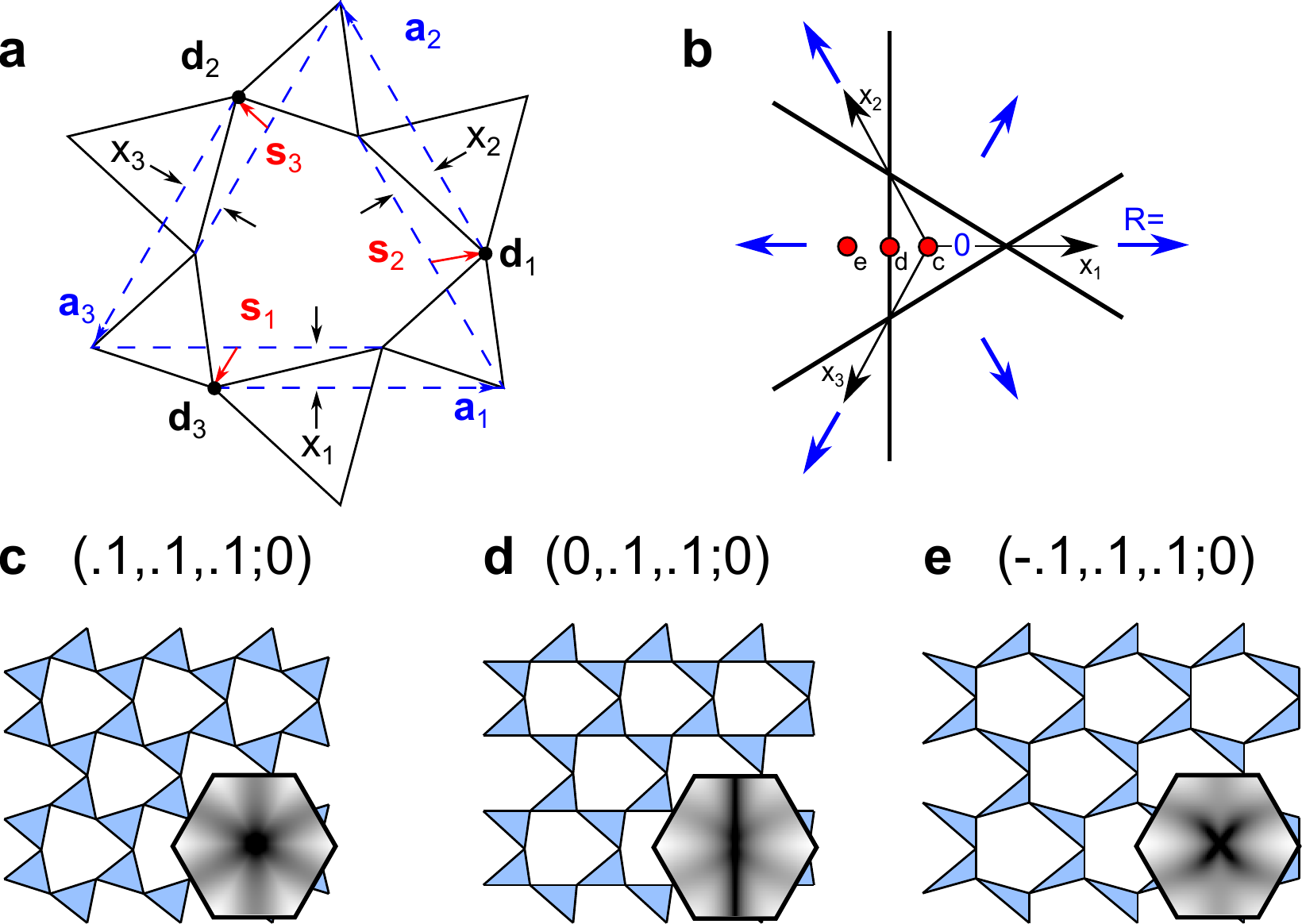}
\caption{{\bf Deformed kagome lattice model. }  {\bf a}
shows our convention for labeling the states.
{\bf b} is a ternary plot of the phase diagram
for fixed $x_1+x_2+x_3>0$.  The phases are labeled
by ${\bf R}$, which is zero in the central phase and a
nearest neighbor lattice vector in the other phases.
{\bf c}, {\bf d} and {\bf e} show representative structures
for ${\bf R}=0$ ({\bf c}) and ${\bf R}\ne 0$ ({\bf e}) and
the transition between them ({\bf d}).  The insets are
density plots of the smallest mode frequency as a function
of ${\bf k}$ in the BZ.  In {\bf c} the gap vanishes only
at ${\bf k}=0$, while in {\bf d} it vanishes on the line
$k_x=0$.  In {\bf e} the gap vanishes only at ${\bf k}=0$,
but has a quadratic dependence in some directions for small ${\bf k}$.}
\label{Fig2}

\end{figure}

We now illustrate the topological boundary modes of a
two-dimensional lattice with the connectivity of the kagome
lattice, but with deformed positions.   The deformed kagome
lattice is specified by its Bravais lattice and basis vectors
for the three atoms per unit cell.  For simplicity, we fix the
Bravais lattice to be hexagonal with primitive vectors ${\bf
a}_{p+1} = (\cos 2\pi p/3,\sin 2\pi p/3)$.   We parameterize
the basis vectors as ${\bf d}_1 = {\bf a}_1/2 + {\bf s}_2$,
${\bf d}_2 = {\bf a}_2/2 - {\bf s}_1$ and ${\bf d}_3 = {\bf
a}_3/2$.  Defining ${\bf s}_3=-{\bf s}_1-{\bf s}_2$,  ${\bf
s}_p$ describe the displacement of ${\bf d}_{p-1}$ relative to
the midpoint of the line along ${\bf a}_p$ that connects its
neighbors at ${\bf d}_{p+1}\pm {\bf a}_{p\mp 1}$ (with $p$
defined mod 3), as indicated in Fig.~2a.  ${\bf s}_p$ are
specified by $6$ parameters with 2 constraints.  A symmetrical
representation is to take ${\bf s}_p = x_p ({\bf a}_{p-1}-{\bf
a}_{p+1}) + y_p {\bf a}_p$ and to use independent variables
$(x_1,x_2,x_3; z)$ with $z = y_1+y_2+y_3$.   The constraints
then determine $y_p = z/3 +  x_{p-1} - x_{p+1}$. $x_p$
describes the buckling of the line of bonds along ${\bf a}_p$,
so that when $x_p=0$ the line of bonds is straight.  $z$
describes the asymmetry in the sizes of the two triangles.
$(0,0,0;0)$ is the undistorted kagome lattice, while
$(x,x,x;0)$ is the twisted kagome lattice, studied in
\cite{SunLub2012}, with twist angle $\theta =
\tan^{-1}(2\sqrt{3}x)$.

It is straightforward to solve for the bulk normal modes of the
periodic lattice. When any of the $x_p$ are zero the gap
vanishes on the line ${\bf k}\cdot {\bf a}_p = 0$ in the BZ.
This line of zero modes is a special property of this model
that follows from the presence of straight lines of bonds along
${\bf a}_p$. When $x_p=0$ the system is at a critical point
separating topologically distinct bulk phases. The phase
diagram features the eight octants specified by the signs of
$x_{1,2,3}$. $(+++)$ and $(---)$ describe states, topologically
equivalent to the twisted kagome lattice, with ${\bf
R}_T = 0$. The remaining 6 octants are states that are
topologically distinct, but are related to each other by $C_6$
rotations. We find
\begin{equation}
{\bf R}_T = \sum_{p=1}^3  {\bf a}_p {\rm sgn} x_p/2
\end{equation}
is independent of $z$. Fig.~2b shows a ternary plot of the
phase diagram  as a function of $x_1,x_2,x_3$ for $z=0$ and a
fixed value of $x_1+x_2+x_3$.   Fig.~2c,d,e show representative
structures for the ${\bf R}_T=0$ phase (Fig.~2c), the ${\bf
R}_T \ne 0$ phase (Fig.~2e), and the transition between them
(Fig.~2d).  The insets show density plots of the lowest
frequency mode, which highlight the gapless point due to the
acoustic mode in Fig.~2c and the gapless line due to states of
self stress in Fig.~2d. In Fig.~2e, the gap vanishes only at
the origin, but the cross arises because  acoustic modes vary
quadratically rather than linearly with ${\bf k}$ along its
axes.  This behavior will be discussed in the next section.

We next examine the boundary modes of the deformed kagome
lattice. Fig.~3 shows a system with periodic boundary
conditions in both directions that has domain walls separating
$(.1,.1,.1;0)$ from $(.1,.1,-.1;0)$. Since there are no broken bonds,
the local count is $\nu_S^L=0$.  On the two domain walls,
equation (\ref{nutcell}) predicts
\begin{equation}
\tilde\nu_T = {\bf G}\cdot ({\bf R}_T^1 -{\bf R}_T^2) = +1 (-1) ,
\end{equation}
for the left (right) domain wall, where ${\bf R}_T^1$ and ${\bf
R}_T^2$ characterize the material to the left and right of the
domain wall, respectively (See fig.~1 in the supplementary
material). Note that the indices of the two domain walls are
opposite in sign so that the total index $\nu$ of equation
(\ref{eq1}) is zero as required. Fig.~2c shows the spectrum of
${\cal H}$ (which has both positive and negative eigenvalues)
as a function of the momentum $k_x$ parallel to the domain
wall. The zero modes of ${\cal H}$ include both the floppy
modes and the states of self-stress. In the vicinity of $k_x=0$
the zero modes on the two domain walls couple and split because
their penetration depth diverges as $k_x\rightarrow 0$. The
eigenvectors for the zero modes at $k_x=\pi$ are indicated in
Fig.~3a by the arrows and the thickened bonds.

\begin{figure}
\includegraphics[width=3.25in]{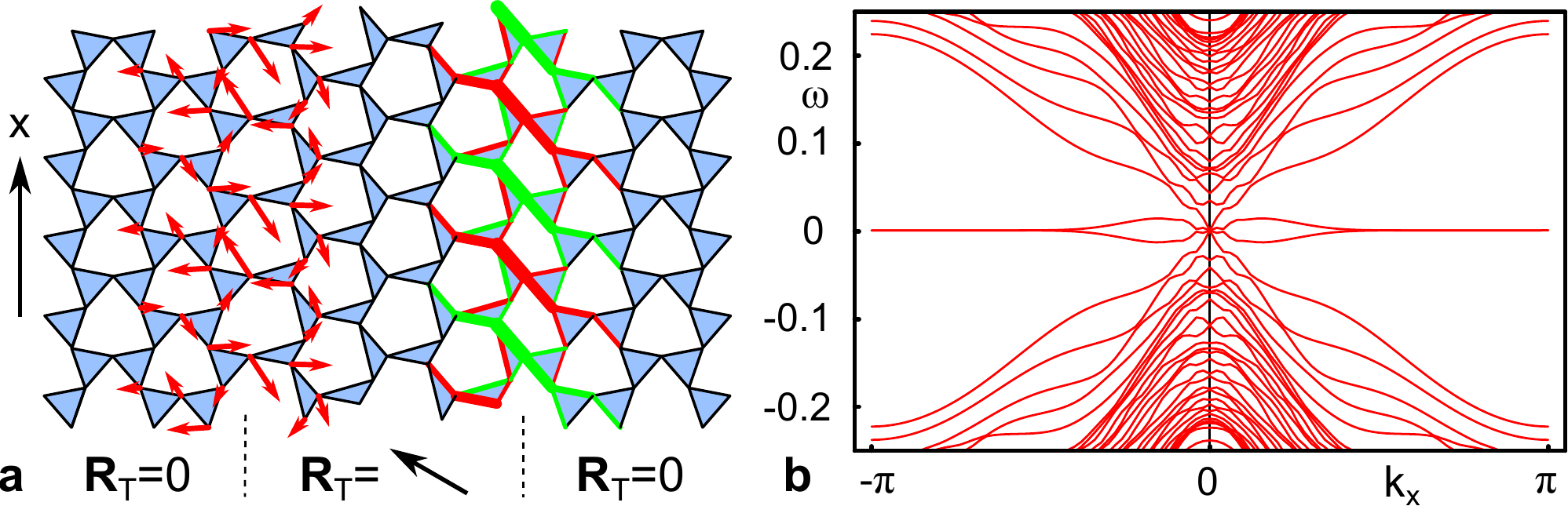}
\caption{{\bf Zero modes at domain walls.}  {\bf a} shows a
lattice with periodic boundary conditions and two domain walls,
the left one between $(.1,.1,.1,0)$ and $(.1,.1,-.1,0)$ with zero modes
and the right one between
$(.1,.1,-.1,0)$ and $(.1,.1,+.1,0)$ with states of self stress.
The zero mode eigenvectors at $k_x=\pi$ are indicated
for the floppy mode (arrows) and the state of self-stress
(red (+) and green (-) thickened bonds).   {\bf b} shows the vibrational
mode dispersion as a function of $k_x$.  }
\label{Fig3}
\end{figure}

Fig.~4a shows a segment of a $(-.05,.05,.05;0)$ lattice with
three different different edges.  For each edge, a unit cell
that accommodates the edge is shown, along with the
corresponding ${\bf R}_L$, from which $\tilde\nu_L$ is
determined.  In the interior, a ``standard" unit cell, with
${\bf r}_0 = 0$ is shown.  Figs.~4b, c, d show the spectrum for
a strip with one edge given by the corresponding edge in
Fig.~4a with free boundary conditions.   The other edge of the
strip is terminated with clamped boundary conditions, so that
the floppy modes are due solely to the free edge.   The number
of zero modes per unit cell agrees with equations
(\ref{nutcell}) and (\ref{nulcell}) for each surface given
${\bf R}_L$, ${\bf R}_T$. The zero modes acquire a finite
frequency when the penetration length of the zero mode
approaches the strip width, which leads to Gaussian ``bumps"
near $k=0$, which will be discussed in the next section. In
Fig.~4d, one of the three zero modes can be identified as a
localized ``rattler", which remains localized on the surface
sites, even for $k\rightarrow 0$.

\begin{figure}
\includegraphics[width=3.25in]{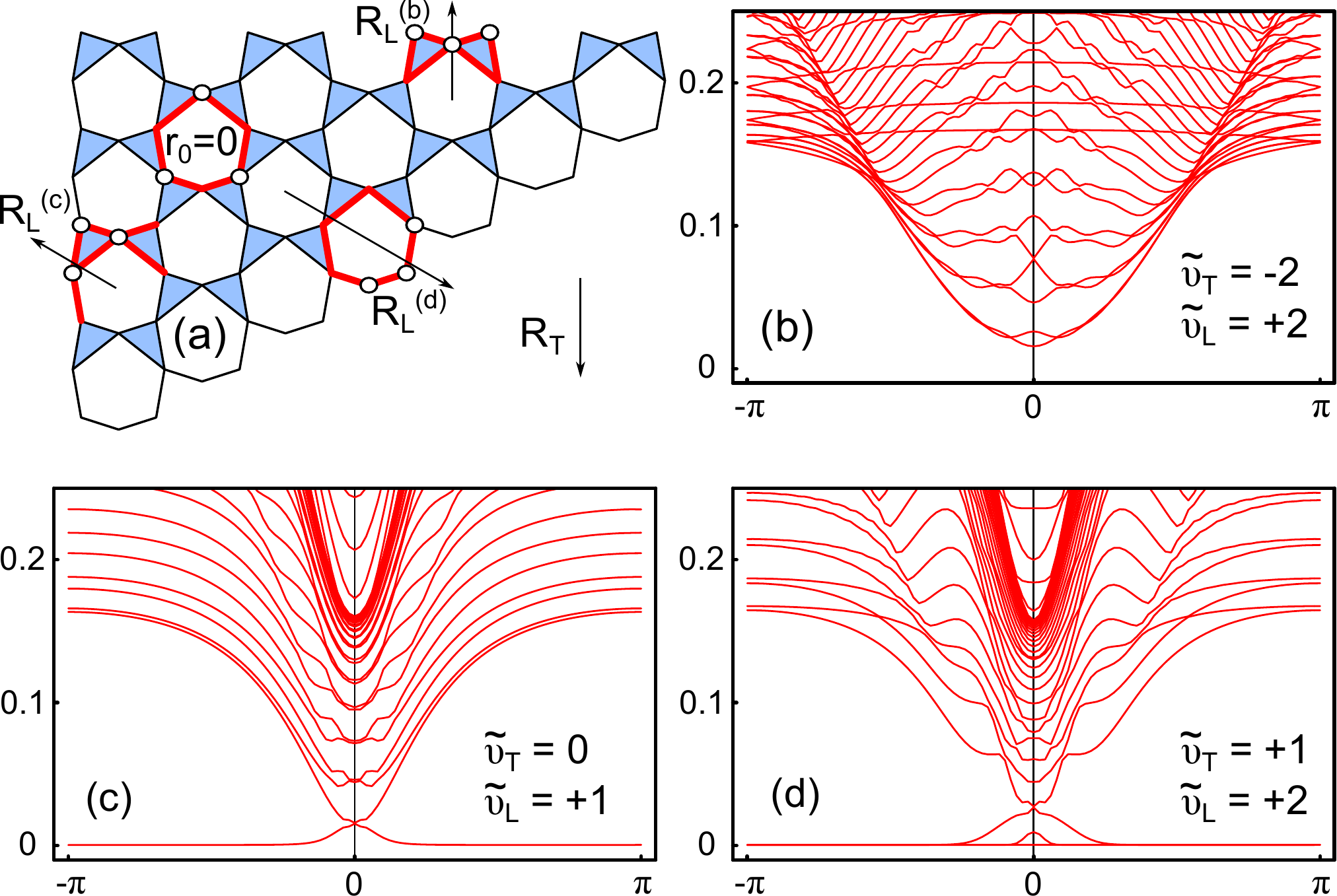}
\caption{{\bf Zero modes at the edge.}
{\bf a} shows a $(-.05,.05,.05,0)$ lattice indicating three edges.
{\bf b}, {\bf c} and {\bf d} show the vibrational mode spectrum computed \
for a strip with one edge as shown in {\bf a} and the other edge
with a clamped boundary condtion.  The zero mode count on each
surface is compared with equations (\ref{eq3},\ref{nutcell},\ref{nulcell}).}
\label{Fig4}
\end{figure}

\section*{Continuum Elasticity Theory}

Unlike electronic spectra, phonon spectra have acoustic modes
whose frequencies vanish as ${\bf k}\rightarrow 0$.  These
excitations along with macroscopic elastic distortions and
long-wavelength surface Rayleigh waves are described by a
continuum elastic energy quadratic in the elastic strain tensor
$u_{ij}$. Elastic energies are restricted to small wavenumber
and cannot by themselves determine topological characteristics,
like those we are considering, that depend on properties across
the BZ.  Nevertheless the elastic energies of our model
isostatic lattices fall into distinct classes depending on the
topological class of the lattice.   For simplicity we focus on
$(x_1,x_2,x_2;0)$ states, where $x_2>0$ is fixed and $x_1$ is
allowed to vary. The elastic energy density $f$ can be written
\begin{equation}
f={K\over 2}[(u_{xx}-a_1 u_{yy})^2 + 2 a_4 u_{xy}^2] .
\label{elastic}
\end{equation}
We find that $a_{1}\propto x_1$ for small $x_1$, while $a_4>0$
and $K$ are positive and smoothly varying near $x_1=0$. Thus,
the ${\bf R}_T=0$ and ${\bf R}_T \neq 0$ sectors are
distinguished by the sign of $a_1$. $f=0$ for shape distortions
with $u_{xx} = a_1 u_{yy}$ and $u_{xy}=0$. When $a_1>0$, the
distortion has a negative Poisson ratio \cite{Lakes1987},
expanding or contracting in orthogonal directions (a feature
shared by the twisted kagome lattice \cite{SunLub2012}); when
$a_1<0$, the distortion has the more usual positive Poisson
ratio. Finally when $a_1=0$, uniaxial compressions along $y$
costs no energy.

Expanding $\det Q^T$ for small $\kv$ provides useful
information about the bulk- and surface-mode structure. To
order $k^3$,
\begin{equation}
\det Q^T = A[k_x^2+ a_1 k_y^2 + i c (k_x^3 - 3 k_x k_y^2)]
+O(k^4),
\label{eq:detQT}
\end{equation}
where $A, c> 0$ for small $x_1$.  $a_1$ is the same as in
equation (\ref{elastic}). Long-wavelength zero modes are
solutions of $\det Q^T = 0$.   The quadratic term, which
corresponds to the elastic theory, equation (\ref{elastic}),
reveals an important difference between the bulk acoustic modes
of ${\bf R}_T= 0$ and ${\bf R}_T \neq 0$. In the former case,
$a_1>0$, $\det Q^T =0$ only at ${\bf k}=0$. For $a_1<0$,
though, to order $k^2$, $\det Q^T = 0$ for $k_x = \pm
\sqrt{|a_1|} k_y$, so the elastic theory predicts lines of
gapless bulk modes. The degeneracy is lifted by the $k^3$ term,
leading to a $k^2$ dispersion along those lines, which can be
seen by the cross in the density map of Fig.~\ref{Fig2}e.

$\det Q^T({\bf k}\rightarrow 0)$ vanishes for complex
wavenumbers associated with zero-frequency Rayleigh surface
waves.   Writing ${\bf k} = k_{\perp} \hat n + k_{||} \hat z
\times \hat n$ for a surface whose outward normal $\hat n$
makes an angle $\theta$ with $\hat x$, there is an $\omega=0$
Rayleigh wave with penetration depth $| {\rm
Im}\,k_{\perp}|^{-1}$ if ${\rm Im}\, k_{\perp} <0$. To order
$k_{||}^2$ there are two solutions,
\begin{equation}
k_{\perp}^\pm=
\frac{\sin \theta \pm i \sqrt{a_1}\cos \theta}{\cos \theta \mp i \sqrt{a_1} \sin \theta} k_{||}
+ \frac{i ( 3 + a_1) d}{2(\cos \theta \pm i \sqrt{a_1} \sin \theta )^3}
k_{||}^2 .
\end{equation}
When $a_1>0$, the linear term is always finite and nonzero, and
${\rm Im}\ k_{\perp}^\pm$ have opposite signs, indicating that
there can be acoustic surface zero modes on all surfaces. These
are the classical Rayleigh waves predicted by the elastic
theory, with penetration depth ${\cal O}(k_{||}^{-1})$. When
$a_1<0$, the linear term in $k_{||}$ is real and ${\rm Im}\,
k_{\perp}^{\pm} \propto k_{||}^2$. The number of long
wavelength surface zero modes depends on the angle of the
surface.  When $|\theta| <\theta_c = \cot^{-1} \sqrt{|a_1|}$,
${\rm Im}\,k_{\perp}^{\pm}$ are both positive, and there are no
acoustic surface zero modes.   The opposite surface,
$|\theta-\pi|<\theta_c$, has two acoustic surface modes.  For
$\theta_c < \theta < \pi - \theta_c$ ${\rm
Im}\,k_{\perp}^{\pm}$ have opposite sign, so there is one mode.
This is consistent with the mode structure in Fig.~\ref{Fig4}:
The ${\cal O}(k_{||}^{-2})$ penetration depth explains the
Gaussian profile of the $k\rightarrow 0$ bumps in the zero
modes, which are due to the finite strip width. In (b) a
$\theta = 0$ surface has no zero modes.  (c) shows a $\theta =
\pi/2 > \theta_c$ surface with one long-wavelength surface zero
mode.  (d) shows the spectrum with $\pi-\theta = \pi/6 <
\theta_c$ with two bumps indicating two deeply penetrating
long-wavelength zero modes in addition to one non-acoustic mode
localized at that surface.

\section*{Conclusions}
We have developed a general theory of topological
phases of isostatic lattices, which explains the boundary zero
modes and connects to the topological band theory of electronic
systems. This points to several directions for future studies.
It will be interesting to study 3D lattice models, as well as
lattices that support point singularities in $\det Q({\bf k})$
analogous to Dirac semimetals.  Finally, it will be interesting
to explore connections with theories of frustrated
magnetism\cite{lawler13}.

Correspondence and requests for materials should be sent to Tom
Lubensky.

\begin{acknowledgements} \
TCL is grateful for the hospitality of the Newton
Institute, where some of this work was carried out.  This work
was supported in part by a Simons Investigator award to CLK
from the Simons Foundation and by the National Science
Foundation under DMR-1104707 (TCL) and DMR-0906175 (CLK).
\end{acknowledgements}

\smallskip\noindent {\bf Author contributions}: CLK and TCL
contributed to the formulation of the problem, theoretical
calculations, and the preparation of the manuscript.

\noindent {\bf Competing financial interests statement}: The
authors have no competing financial interests to declare.

\section*{SUPPLEMENTARY ONLINE MATERIALS}

\subsection*{Proof of Index Theorem}

In this appendix we provide details of the proof of the index
theorem discussed in the text.   Our starting point is equation
(\ref{eq4}), which describes the zero-mode count in a region
$S$ of a larger system.  Using the fact that $\{H,\tau^z\} =
[\rho_S(\hat{\bf r}),\tau^z] = 0$ it is straightforward to
check that equations (\ref{eq3}-\ref{nul}) imply that
\begin{equation}
\nu^S_T =  {1\over 2}\lim_{\epsilon\rightarrow 0}{\rm Tr} \left[ \tau^z {1\over{{\cal H}+i\epsilon}} [ \rho^S(\hat {\bf r}), {\cal H}]\right].
\end{equation}
Since $[\rho^S,{\cal H}]=0$ for $\rho_S = 1$, and ${\cal H}$
has a finite range $a$, $\nu_T^S$ comes only from the boundary
of region $S$ where $\rho^S({\bf r})$ varies.  If we assume
that the boundary region is gapped and that $\rho({\bf r})$
varies slowly on the scale $L\gg a$, then we safely take
$\epsilon$ to zero and expand to leading order in $a/L$.  Since
$[\rho^S(\hat{\bf r}),{\cal H}]_{\alpha\beta} = {\cal
H}_{\alpha\beta} (\rho^S({\bf r}_\alpha)-\rho^S({\bf r}_\beta))
\sim {\cal H}_{\alpha\beta}({\bf r}_\alpha-{\bf r}_\beta) \cdot
\nabla\rho({\bf r}_\beta)$, we may write
\begin{equation}
\nu^S_T = {1\over 2}{\rm Tr} \left[\tau^z \nabla\rho(\hat {\bf r}) \cdot {\cal H}^{-1} [\hat {\bf r},{\cal H}] \right].
\end{equation}

We next suppose that in the boundary region the lattice is
periodic, so that the trace may be evaluated in a basis of
plane waves:
\begin{equation}
|{\bf k},a\rangle = {1\over \sqrt{N}}\sum_{{\bf R}}\exp i {\bf k}\cdot({\bf R} + {\bf d}_a)|{\bf R} + {\bf d}_a \rangle,
\end{equation}
where ${\bf R}$ is a Bravais lattice vector in a system with
periodic boundary conditions and $N$ unit cells.  ${\bf d}_a$
are basis vectors for the $d n_s + n_b$ sites and bonds per
unit cell.  The phases are chosen such that the position
operator is $\hat {\bf r} \sim i \nabla_{\bf k}$.  In this
basis, the Bloch Hamiltonian ${\cal H}({\bf k})$ is a $d n_s +
n_b$ square matrix with off diagonal blocks $Q({\bf k})$ and
$Q^\dagger({\bf k})$, where
\begin{equation}
Q_{ab}({\bf k}) = \langle {\bf k},a|Q| {\bf k},b\rangle.
\end{equation}
$\nu_T^S$ then has the form
\begin{equation}
\nu^S_T = \int_{\partial S} {d^{d-1}S} \, {\bf P}_T \cdot \hat n
\label{nutsupp}
\end{equation}
where the integral is over the boundary of $S$ with inward
normal $\hat n$, and
\begin{equation}
{\bf P}_T = \int_{BZ} {d^d{\bf k}\over{(2\pi)^d}} {\rm Im}\ {\rm Tr}[Q^{-1} \nabla_{\bf k} Q].
\label{p}
\end{equation}

It is useful, to write
\begin{equation}
{\rm Im}{\rm Tr}[Q^{-1}\nabla_{\bf k} Q] = \nabla_{\bf k} {\rm Im} \log \det Q.
\end{equation}
It is then straightforward to show that
\begin{equation}
\det Q({\bf k} + {\bf G}) = \det Q({\bf k}) \exp[-i {\bf G} \cdot {\bf r}_0],
\end{equation}
 where
\begin{equation}
{\bf r}_0 = d\sum_{{\rm sites}\ i} {\bf d}_i - \sum_{{\rm bonds}\ m} {\bf d}_m.
\end{equation}
For a general lattice, ${\bf r}_0$ is non zero.   However, if
the coordination number of site $i$ is $z_i$ then ${\bf r}_0 =
\sum_i (d-z_i/2) {\bf d}_i + {\bf R}$, where ${\bf R}$ is a
Bravais lattice vector.   Thus, for an isostatic lattice with
uniform coordination $z=2d$, ${\bf r}_0$ is a Bravais lattice
vector, and it is always possible to shift ${\bf d}_m$ by
lattice vectors to make ${\bf r}_0 = 0$.   In the text of the
paper, we assumed ${\bf r}_0=0$.   Here we will keep it
general, and show that while ${\bf r}_0$ affects $\nu^S_T$, its
effect is canceled by a compensating term in $\nu^S_L$.

For the general case, let us write $\det Q({\bf k}) =
q_0({\bf k}) \exp[-i {\bf k} \cdot {\bf r}_0] $, where
$q_0({\bf k}) = q_0({\bf k}+ {\bf G})$ is periodic in the BZ.
Equation (\ref{p}) then involves two pieces:
\begin{equation}
{\bf P}_T = {1\over{V_{\rm cell}}} \left[-{\bf r}_0 +  {\bf R}_T\right].
\label{psupp}
\end{equation}
Here ${\bf R}_T$ is a Bravias lattice vector describing the
winding numbers of the phase of $q_0({\bf k})$ around the
cycles of the BZ.  It may be written ${\bf R}_T = \sum_i n_i
{\bf a}_i$ with
\begin{equation}
n_i = {1\over{2\pi i}}\int_{C_i}  d{\bf k} \cdot  \nabla_{\bf k} \log q_0({\bf k})
\label{nia}
\end{equation}
where as in the text, we assume that for a given cycle $C_i$ of
the BZ the winding number is path indpendent.

\setcounter{figure}{0}
\begin{figure}
\includegraphics[width=3.in]{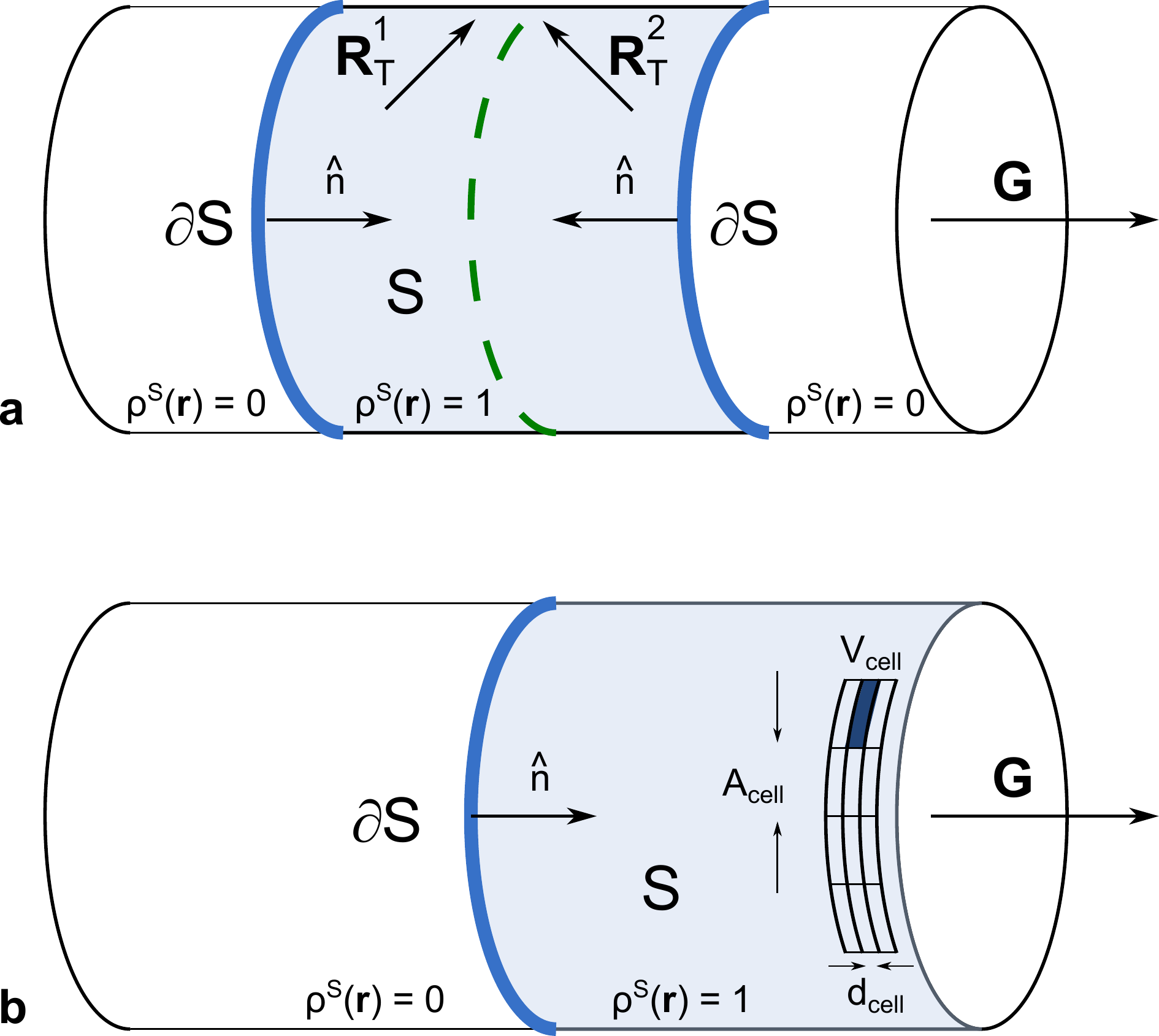}
\caption{{\bf Evaluating the zero mode count.}  {\bf a} Cylindrical geometry for evaluating the zero mode count for a domain wall between ${\bf R}_T^1$ and ${\bf R}_T^2$, indicated by the dashed line.  {\bf b} Cylindrical geometry for evaluating the zero mode count for a surface indexed by reciprocal lattice vector ${\bf G}$.  The region $S$ covers half the cylinder.  The boundary $\partial S$ is deep in the interior.  {\bf b} also shows our notation for the surface unit cell.}
\label{FigS1}
\end{figure}

\subsubsection*{Application to zero modes at a domain wall}

To determine the zero mode count at a domain wall between
topological states ${\bf R}_T^1$ and ${\bf R}_T^2$, we consider
a cylinder perpendicular to the domain wall (or a similar
construction for $d$ dimensions). We expect the zero mode count
to be proportional to the ``area" $A$ (or length in 2D) of the
domain wall. We will, therefore, be interested in the zero mode
count per unit cell, $\nu^S/N_{\rm cell}$, where $N_{\rm cell}
= A/A_{\rm cell}$, and $A_{\rm cell} = V_{\rm cell}/d_{\rm
cell}$ is the projected area of the surface unit cell, which
can be expressed in terms of the volume of the bulk unit cell
$V_{\rm cell}$ and the distance $d_{\rm cell}$ between Bragg
planes. Referring to Supplementary Fig. 1a, we use equation
(\ref{psupp}) to evaluate equation (\ref{nutsupp}) away from
the domain wall to give $\nu_T = (A/V_{\rm cell}) ({\bf R}_T^1
- {\bf R}_T^2)\cdot \hat n$, where $\hat n$ is the unit vector
pointing to the right.   The zero mode count per unit cell can
be expressed in terms of the reciprocal lattice vector ${\bf G}
= 2\pi \hat n/d_{\rm cell}$ that indexes the domain wall as
\begin{equation}
\nu_T^S/N_{\rm cell} = {\bf G}\cdot ({\bf R}_T^1 - {\bf R}_T^2)/2\pi.
\end{equation}

\subsubsection*{Application to zero modes at the edge}

We next determine the number of zero modes
localized on a surface (or edge in 2d) indexed by a reciprocal
lattice vector ${\bf G}$.  Consider a cylinder with axis
perpendicular to ${\bf G}$ and define the region $S$ to be the
points nearest to one end of the cylinder, as shown in Supplementary Fig. 1b.
A similar
construction can be used to count the zero modes on a surface
in $d$ dimensions.

$\nu^S_T$ is determined by evaluating equation (\ref{nut}) deep in the
bulk of the cylinder where the lattice is periodic.   From equation
(\ref{p}) we may write
\begin{equation}
\nu^S_T/N_{\rm cell} = {\bf G} \cdot ({\bf R}_T- {\bf r}_0)/2\pi.
\end{equation}

The local count, $\nu^S_L$, depends on the details of the
termination at the surface and is given by the macroscopic
``surface charge" that arises when positive charges $+d$ are
placed on the sites and negative charges $-1$ are placed on the
bonds.   As discussed in the text, it can be determined by
evaluating the dipole moment of a unit cell with site and bond
vectors $\tilde {\bf d}_a$ that is defined so that the surface
can be accomodated with no left over sites or bonds.   This
unit cell is in general different from the unit cell used to
compute $\nu^S_T$, and its dipole moment is in general not
quantized.   However, since the difference is due to a
redefinition of which bond is associated with which unit cell,
the dipole moment differs from ${\bf r}_0$ by a Bravais lattice
vector,
\begin{equation}
{\bf R}_L = d \sum_{{\rm sites }\ i} \tilde{\bf d}_i - \sum_{{\rm bonds }\ m} \tilde {\bf d}_m - {\bf r}_0.
\end{equation}
It follows that the local count may be written
\begin{equation}
\nu^S_L/N_{\rm cell} = {\bf G}\cdot ({\bf R}^L + {\bf r}_0)/2\pi.
\end{equation}

The total zero mode count on the edge is then
\begin{equation}
\nu^S/N_{\rm cell} = {\bf G} \cdot ({\bf R}^L + {\bf R}^T)/2\pi.
\label{nus}
\end{equation}
It can be seen that the dependence on ${\bf r}_0$, which
depends on the arbitrary unit cell used to define $\nu^S_T$
cancels.

\end{document}